\documentclass[aps, twocolumn, prapplied, superscriptaddress]{revtex4}

\usepackage{graphicx,float}
\usepackage{epsfig}
\usepackage{epstopdf}

\usepackage{amsmath}
\usepackage{amssymb}
\usepackage{bm}
\usepackage{bbm}
\usepackage{dsfont}
\usepackage{bbold}
\usepackage{multirow}
\usepackage{color}
\usepackage[usenames,dvipsnames]{xcolor}
\usepackage[colorlinks=true,citecolor=blue,urlcolor=blue]{hyperref}

\usepackage{hyperref}

\hypersetup{colorlinks=true,linkcolor=blue,citecolor=blue,urlcolor=blue}

\newcommand{\be}{\begin{equation}}
\newcommand{\ee}{\end{equation}}

\newcommand{\sket}[1]{{\ensuremath{\lvert#1\rangle}}}
\newcommand{\lket}[1]{{\ensuremath{\left\lvert#1\right\rangle}}}
\newcommand{\ket}[1]{\if@display\lket{#1}\else\sket{#1}\fi}

\newcommand{\sbra}[1]{{\ensuremath{\langle#1\rvert}}}
\newcommand{\lbra}[1]{{\ensuremath{\left\langle#1\right\rvert}}}
\newcommand{\bra}[1]{\if@display\lbra{#1}\else\sbra{#1}\fi}

\newcommand{\sbraket}[2]{{\ensuremath{\langle#1\rvert#2\rangle}}}
\newcommand{\lbraket}[2]{{\ensuremath{\left\langle#1\!\left\rvert\vphantom{#1}#2\right.\!\right\rangle}}}
\newcommand{\braket}[2]{\if@display\lbraket{#1}{#2}\else\sbraket{#1}{#2}\fi}

\newcommand{\sketbra}[2]{{\ensuremath{\lvert #1\rangle\!\langle #2\rvert}}}
\newcommand{\lketbra}[2]{{\ensuremath{\left\lvert #1\right\rangle\!\!\left\langle #2\right\rvert}}}
\newcommand{\ketbra}[2]{\if@display\lketbra{#1}{#2}\else\sketbra{#1}{#2}\fi}

\begin{document}


\title{Few-mode-fiber technology fine-tunes losses of quantum communication systems}


\author{A. Alarc\'{o}n}
\affiliation{Institutionen f\"or Systemteknik, Link\"opings Universitet, 581 83 Link\"oping, Sweden}

\author{J. Argillander}
\affiliation{Institutionen f\"or Systemteknik, Link\"opings Universitet, 581 83 Link\"oping, Sweden}

\author{G. Lima}
\affiliation{Departamento de F\'{\i}sica, Universidad de Concepci\'on, 160-C Concepci\'on, Chile}
\affiliation{Millennium Institute for Research in Optics, Universidad de Concepci\'on, 160-C Concepci\'on, Chile}

\author{G. B.~Xavier}
\email{guilherme.b.xavier@liu.se}
\affiliation{Institutionen f\"or Systemteknik, Link\"opings Universitet, 581 83 Link\"oping, Sweden}


\begin{abstract}
A natural choice for quantum communication is to use the relative phase between two paths of a single-photon for information encoding. This method was nevertheless quickly identified as impractical over long distances and thus a modification based on single-photon time-bins has then become widely adopted. It however, introduces a fundamental loss, which increases with the dimension and that limits its application over long distances. Here, we are able to solve this long-standing hurdle by employing a few-mode fiber space-division multiplexing platform working with orbital angular momentum modes. In our scheme, we maintain the practicability provided by the time-bin scheme, while the quantum states are transmitted through a few-mode fiber in a configuration that does not introduce post-selection losses. We experimentally demonstrate our proposal by successfully transmitting phase-encoded single-photon states for quantum cryptography over 500 m of few-mode fiber, showing the feasibility of our scheme.
\end{abstract}




\maketitle


\section{Introduction}

Quantum Communication (QC) is one of the main pillars of the applied field of Quantum Technologies \cite{Acin_2018}, which deals with information processing tasks that rely on individual and entangled quantum systems. QC includes many applications such as quantum cryptography \cite{Xu_2020, Pirandola_2020}, quantum bit commitment \cite{Lunghi_2013} and quantum secret sharing \cite{Pinnell_2020}. Although traditionally many experiments have been performed while resorting to polarization-encoded quantum states \cite{Xavier_2009, Grunenfelder_2018}, time-bin-based phase-coding quantum cryptography has always been regarded as the optimal choice for optical fiber communication links due to strong robustness to environmental disturbances \cite{Bennett_1992, Gisin_2002, Townsend_1993, Gobby_2004, Wang_2008, Yuan_2008, Wang_2012, Rubenok_2013, Yang_2013, Islam_2017a, Boaron_2018}. The phase-coding quantum cryptography protocol was initially discussed considering the relative phase between two different single-photon spatial modes defining a long Mach-Zehnder interferometer  \cite{Bennett_1992, Gisin_2002}. The main limitation of this original scheme was that the interferometer length needed to be as long as the physical separation between the communicating parties (Alice and Bob), thus subjected to strong environmental phase disturbances. More recently, the technology to stabilise long Mach-Zehnder interferometers supporting the propagation of single photons with short coherence length has become available \cite{Xavier_2012, Cuevas_2013, Carvacho_2015}, but it nevertheless adds to the experimental complexity.  

The time-bin based configuration was proposed to solve this issue \cite{Bennett_1992, Gisin_2002}. It trades the long single Mach-Zehnder interferometer of the previous scheme with two unbalanced and short Mach-Zehnder interferometers (UMZIs), one located at Alice and the other within Bob's station. Both interferometers are connected with a single optical fiber comprising the communication channel linking Alice and Bob. The key point is that the two long parallel paths of the original phase-encoding scheme are traded for two temporally-separate time-bins that travel through the same path. This guarantees intrinsic phase stability, since any phase disturbance needs to occur on a time scale shorter than the separation between the two time-bins. Only the short UMZIs need to be stabilized, a task considerably easier than actively compensating phase drifts on a long interferometer. The trade-off is an intrinsic loss of 50\% at the detection stage due a post-selection of non-interfering time-bins. Even more dramatic, this intrinsic loss grows with the dimension $d$ of the encoded system as $(d-1)/d$, having a larger impact on high-dimensional time-bin QC systems \cite{Islam_2017b}. In spite of this, time-bin has been extremely popular since it is much more practical to stabilise two short interferometers instead of a very long one \cite{Gisin_2002}.

In this work, we adopt modern space-division multiplexing (SDM) fiber optics technology \cite{Richardson_2013,Xavier_2020} to solve the issue of the irreversible loss at the detection stage of the time-bin scheme, thus providing a pathway towards efficient fiber-based phase-coding quantum communication. Specifically, in our scheme we use a commercial SDM fiber that supports a few linearly polarised (LP) spatial modes, the so-called few-mode fiber (FMF). Similar to the time-bin configuration, our scheme is based on a single (few-mode) fiber interconnecting two local interferometers for the state generation and detection. These states are encoded as a superposition of the LP optical modes supported by the FMF. The key component that allows this implementation is the mode-selective photonic lantern, which takes $N$ input single-mode fibers, and maps each one-to-one to $N$ linearly polarised modes of the FMF fiber \cite{Birks_2015}. As there is no difference in time between the different interferometers' paths, no temporal post-selection is needed, and thus there is no irreversible loss in the detection stage. Last, we highlight that our scheme allows the generation and measurement of light modes carrying orbital angular momentum (OAM) in an all-fiber platform, thus dispensing complex mode multiplexers and sorters based on bulk optics \textcolor{red}{\cite{Berkhout_2010, Lavery_2012, Huang_2015, Wen_2020, Cao_2020, Liu_2020}}. The all-in-fiber generation of OAM light modes is not only useful for classical and quantum communication. The compactness that can be achieved for our OAM source/detection modules can certainly find applications in biophysics \cite{Padg11, Grie03, Fran08}, metrology \cite{Damb13nc}, and astronomy \cite{Tamb11} for instance.

 \section{Experiment}
 
\begin{figure}[ht!]
\centering
\includegraphics[width=0.82\columnwidth]{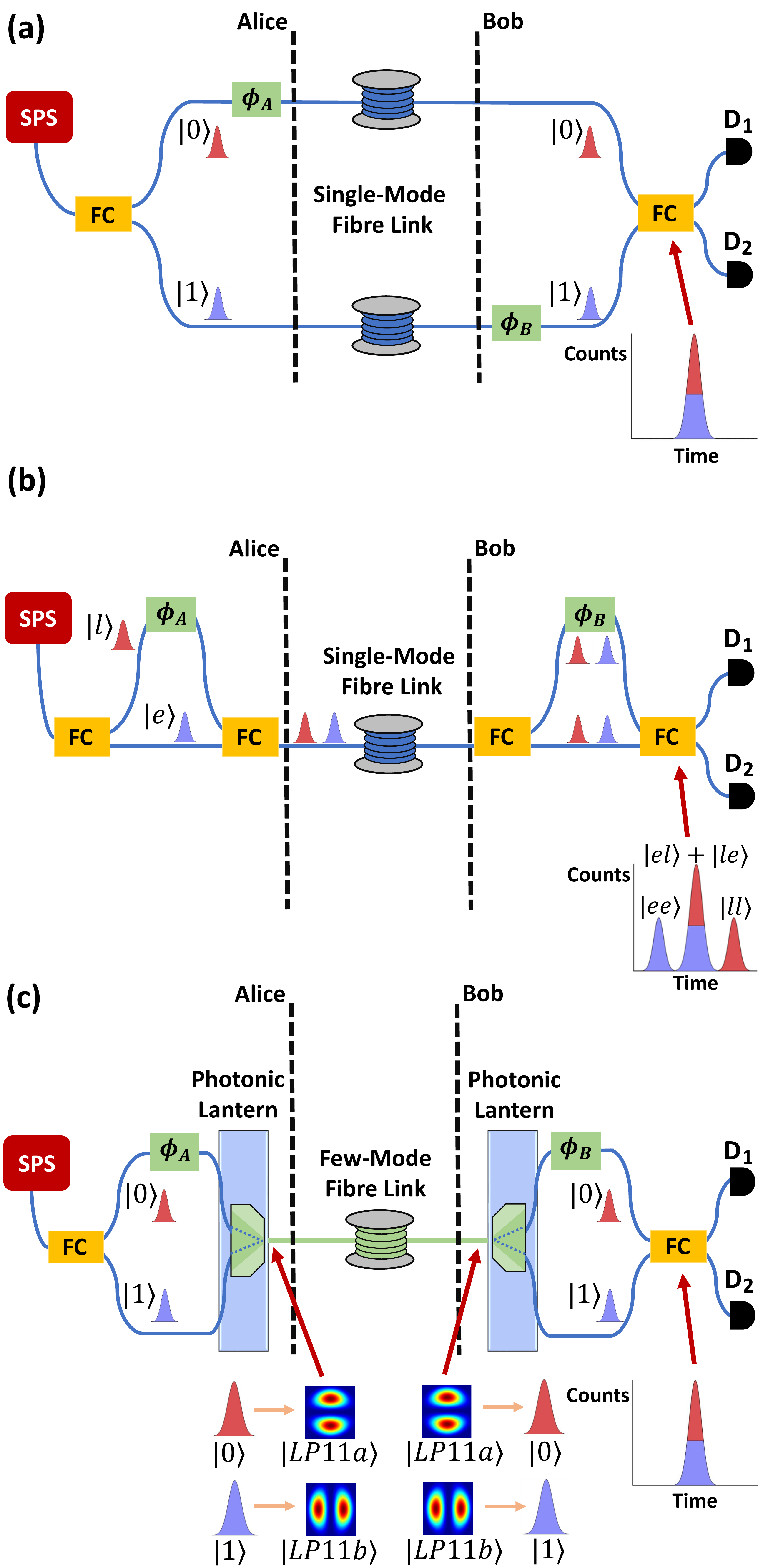}
\caption{Phase-coding quantum cryptography. a) Original scheme based on a long Mach-Zehnder interferometer that must be phase-stabilised. Phase modulators are used for preparing and measuring the required states. The inset shows that photons transmitted through paths $\ket{0}$ and $\ket{1}$ will arrive in the same detection time window. b) Typical time-bin based scheme, relying on a single fiber interconnecting two asymmetrical short interferometers (easier to stabilise). Temporal post-selection is required. Photons will arrive in three different time-bins (see inset). In order to extract a secret key the ones arriving too early or too late must be discarded, thus,  drastically reducing the communication rate by $50\%$. c) Our setup resorting to few-mode fiber technology. It borrows concepts of the robust time-bin configuration: two short interferometers interconnected through a single (few-mode) fiber. Photons travelling through paths $\ket{0}$ and $\ket{1}$ are mapped to orthogonal optical modes supported by the fiber using a photonic lantern, and demultiplexed in an inverse fashion by another lantern. Since the local interferometers are now symmetrical, these photons will again arrive in the same time bin and no temporal post-selection is needed.} \label{Fig1}
\end{figure}

In the original phase-encoded scheme Alice has a single-photon source (SPS) producing time-localised single-photons, whose output is split into two paths through a 50:50 bidirectional fiber coupler (FC). A phase modulator $\phi_A$ is placed in one of the arms, thus producing quantum states given by $|\psi \rangle =  (1/\sqrt{2})\left( |0\rangle + e^{i\phi_A } |1\rangle \right)$, where $|0\rangle$ and $|1\rangle$ correspond to the upper and lower path modes respectively (Fig. \ref{Fig1}a). Both paths are connected to Bob, who also has another phase modulator ($\phi_B$) allowing him to choose between the two required measurement bases. The measurement procedure is concluded with both paths superposed on another 50:50 FC, followed by a single-photon detector placed in each outcome mode. In this case, the single-photon detection probability at D$_1$ (D$_2$) is proportional to $\mathrm{cos}^2(\phi_A - \phi_B)$ [$\mathrm{sin}^2(\phi_A - \phi_B)$], which is a requirement for quantum key distribution for instance \cite{Gisin_2002}.

In the time-bin modification (Fig. \ref{Fig1}b) Alice first splits an attenuated optical pulse over an \textit{early} (\textit{e}) and a \textit{late} (\textit{l}) time-bin with an unbalanced Mach-Zehnder interferometer (UMZI), where the imbalance must be greater than the pulse length. Phase modulator $\phi_A$ is placed inside the interferometer encoding the state $|\psi \rangle =  (1/\sqrt{2})\left( |e\rangle + e^{i\phi_A} |l\rangle \right)$, which propagates towards Bob through an optical fiber. Bob possesses an identical UMZI with phase modulator $\phi_B$ placed in the long arm, which is also used to choose his measurement projection. At the outputs of Bob's UMZI, a temporal post-selection procedure is used to only detect photons that took the $|e\rangle |l\rangle$ or $|l\rangle |e\rangle$ path combinations. These instances are indistinguishable in principle, and thus display a sinusoidal interference pattern that depends on $\phi_A - \phi_B$, which is the same as in the original scheme (Fig. \ref{Fig1}a). The other two possibilities that the photons may take ($|e\rangle |e\rangle$ and $|l\rangle |l\rangle$) are distinguishable, do not interfere and are thus discarded with temporal post-selection (Inset Fig. \ref{Fig1}b). This represents a loss by a factor two at the detector stage, which is the trade-off when employing only one fiber for connecting the communicating parties.

In our experimental demonstration (Fig. \ref{Fig1}c), the single-photon source (SPS) consists of a continuous wave distributed feedback telecom semiconductor laser (Exfo OS-DFB) operating at 1546 nm. The signal is split into two paths ($|0\rangle $ and $|1\rangle$) using a 50:50 FC at Alice's station. A variable optical attenuator (ATT) placed before the fiber coupler is used to change the power intensity to the single-photon level. A lithium niobate pigtailed telecom phase modulator (Thorlabs LN65S-FC) is placed on one path ($\phi_A$). The phase modulator is driven by an electrical signal from a function generator. The two paths are then connected to a commercial mode-selective photonic lantern (Phoenix Photonics 3PLS-GI-15). It allows for multiplexing information into the spatial modes supported by a FMF through the implementation of the following mode mapping: $|1\rangle \to |\textrm{LP}_{11a}\rangle$ ; $|2\rangle \to |\textrm{LP}_{11b}\rangle$. polarization controllers (PC), not shown for simplicity, are placed at each path $|0\rangle $ and $|1\rangle$ to ensure the polarization state of each input mode is the same. We also employ a variable attenuator in tandem with a 90:10 fiber coupler before each input to the photonic lantern in order to ensure each single-mode input has the same optical intensity, thus generating the intended state superpositions.

The FMF link following the lantern consists of either a direct back-to-back connection with only a 10 m long FMF manual polarization controller (MPC), or an added spool of FMF with a length of 500 m, with a measured total loss of 1.2 dB, including two home-made fiber connectors. The MPC is employed to optimize mode demultiplexing at Bob's lantern. The FMFs used in this experiment are commercially available graded-index telecom fiber  (OFS 80730), with a loss coefficient of less than 0.22 dB/km as specified by the manufacturer. The detection stage consists of another photonic lantern that is now used as a demultiplexer, where the inverse mapping is performed. The lantern outputs are the single-mode path-states $\ket{0}$ and $\ket{1}$, which are then recombined on another 50:50 fiber coupler. The measurement basis implemented is defined by a second phase modulator ($\phi_B$). Standard MPCs in each arm are also employed to align the photon polarization state such that in the final interferometer there is no path-information available \cite{QErasure,DecQErasure}, which would compromise the visibility of the observed interference. Following the final beamsplitter, we place InGaAs single-photon counting modules (IdQ id210) operating with a gate width of 2.5 ns, internal trigger rate of 1 MHz, overall detection efficiency of 10\%, and dark count probability per gate of $2.4 \times 10^{-6}$. An in-fiber polariser is placed before each detector. Overall, we obtain -14.6 and -16.2 dB of extinction rate when measuring the outputs $|1\rangle$ and $|2\rangle$ respectively for the opposing inputs ($|2\rangle$ and $|1\rangle$). The insertion loss given by our commercial lanterns is 6.5 dB.

\section {Results}

We focus on the generation of the states that are used for phase-encoding BB84 QKD \cite{Gisin_2002}. These are based on sets of orthogonal states divided between two mutually unbiased basis (MUBs) \cite{Mafu_2013}. In our case, these states are based on coherent superpositions of LP$_{11}$ modes:

\begin{equation}
    |\mathrm{LP}_{\pm}\rangle = \frac{1}{\sqrt{2}}(|\mathrm{LP}_{11a}\rangle \pm |\mathrm{LP}_{11b}\rangle).
\end{equation}
\begin{equation}
    |\mathrm{OAM}_{\pm}\rangle = \frac{1}{\sqrt{2}}(|\mathrm{LP}_{11a}\rangle \pm i|\mathrm{LP}_{11b}\rangle).
\end{equation}

\begin{figure}[ht!]
\centering\includegraphics[width=\columnwidth]{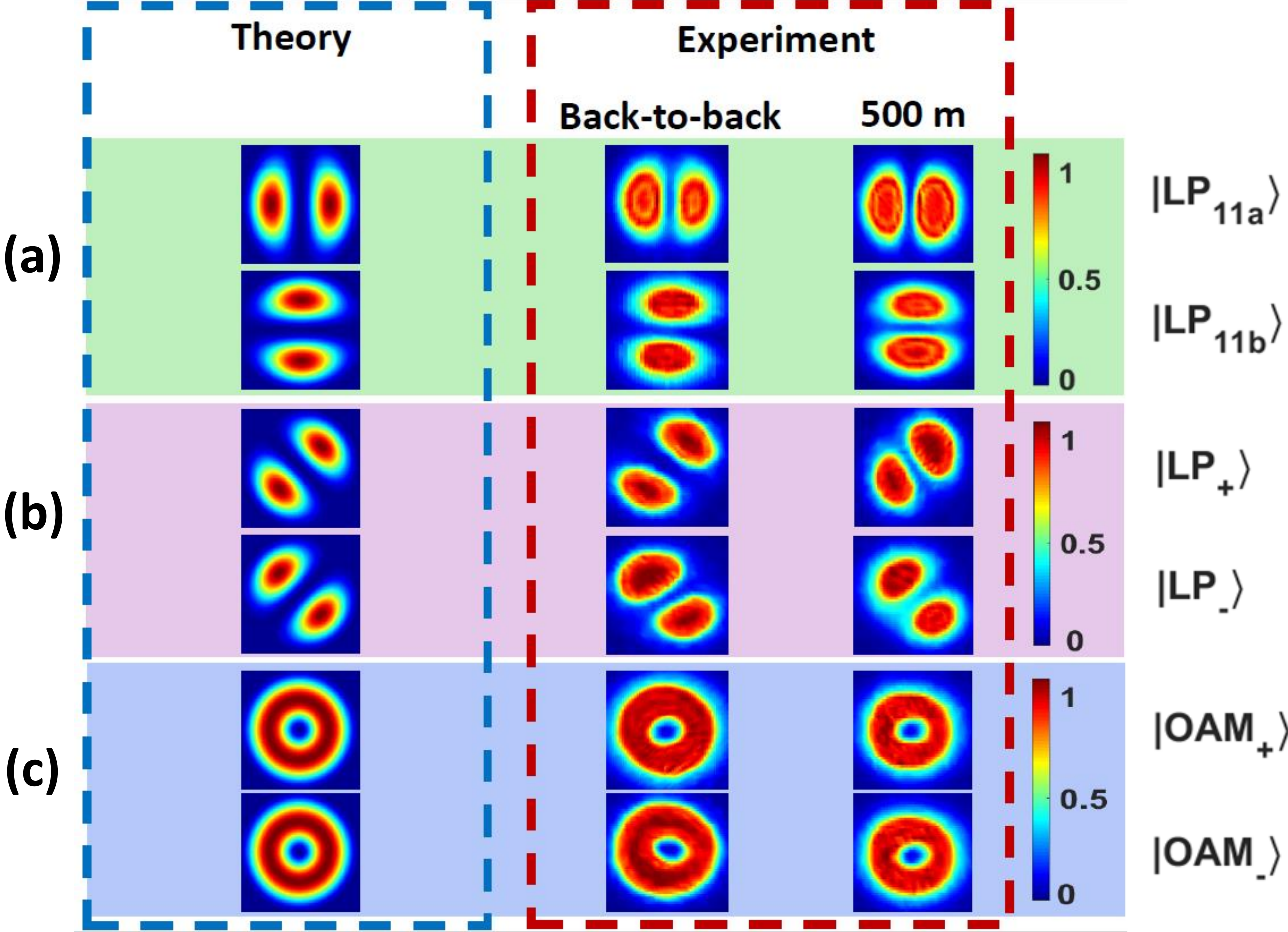}
\caption{Spatial intensity profiles of the output of the FMF as measured by an InGaAs CCD camera. a) Theoretical spatial profiles of the LP$_{11a}$ and LP$_{11b}$ modes and experimental results in the back-to-back case and  after an extra 500 m of FMF. b) and c) Theoretical and experimental spatial profiles (back-to-back and 500 m) associated to the $|\mathrm{LP}_{\pm}\rangle$ and $|\mathrm{OAM}_{\pm}\rangle$ states respectively.
  \label{Fig2} }
\end{figure}

The states $|\mathrm{OAM}_{\pm}\rangle$ are associated to the well known Laguerre-Gaussian beams carrying orbital angular momentum (OAM) \cite{Li_2019}. Figure \ref{Fig2}a shows the theoretical transverse intensity profiles of the $|\mathrm{LP}_{11a}\rangle$ and $|\mathrm{LP}_{11b}\rangle$ modes as well as experimental results in the back-to-back case and after 500 m of propagation, obtained with a linear polariser and an infrared CCD camera placed after the FMF link. We also measured the intensity profiles associated to the states $|\mathrm{LP}_{\pm}\rangle$ and $|\mathrm{OAM}_{\pm}\rangle$, which are shown respectively in Figs. \ref{Fig2}b and \ref{Fig2}c. Here, each profile is prepared by setting the appropriate phase $\phi_A = \{ 0 , \pi , \pi/2 , 3\pi /2\}$ through a slow driving-signal (0.2 Hz) applied to the phase modulator, thus sequentially preparing them. The driving voltage of the phase modulator was previously calibrated. To obtain the intensity profiles of the $|\mathrm{LP}_{11a}\rangle$ and $|\mathrm{LP}_{11b}\rangle$ modes, each input arm was individually blocked. For all these measurements, we worked with the source at the classical regime bypassing the variable optical attenuator. These initial results show that the photonic lantern can be used to prepare the coherent quantum superpositions of the spatial states $|\mathrm{LP}_{11a}\rangle$ and $|\mathrm{LP}_{11b}\rangle$, and also show that these superpositions suffer little degradation following propagation after an extra 500 m, indicating longer distances may be feasible with the current setup.

\begin{figure}[h!]
\centering\includegraphics[width=\columnwidth]{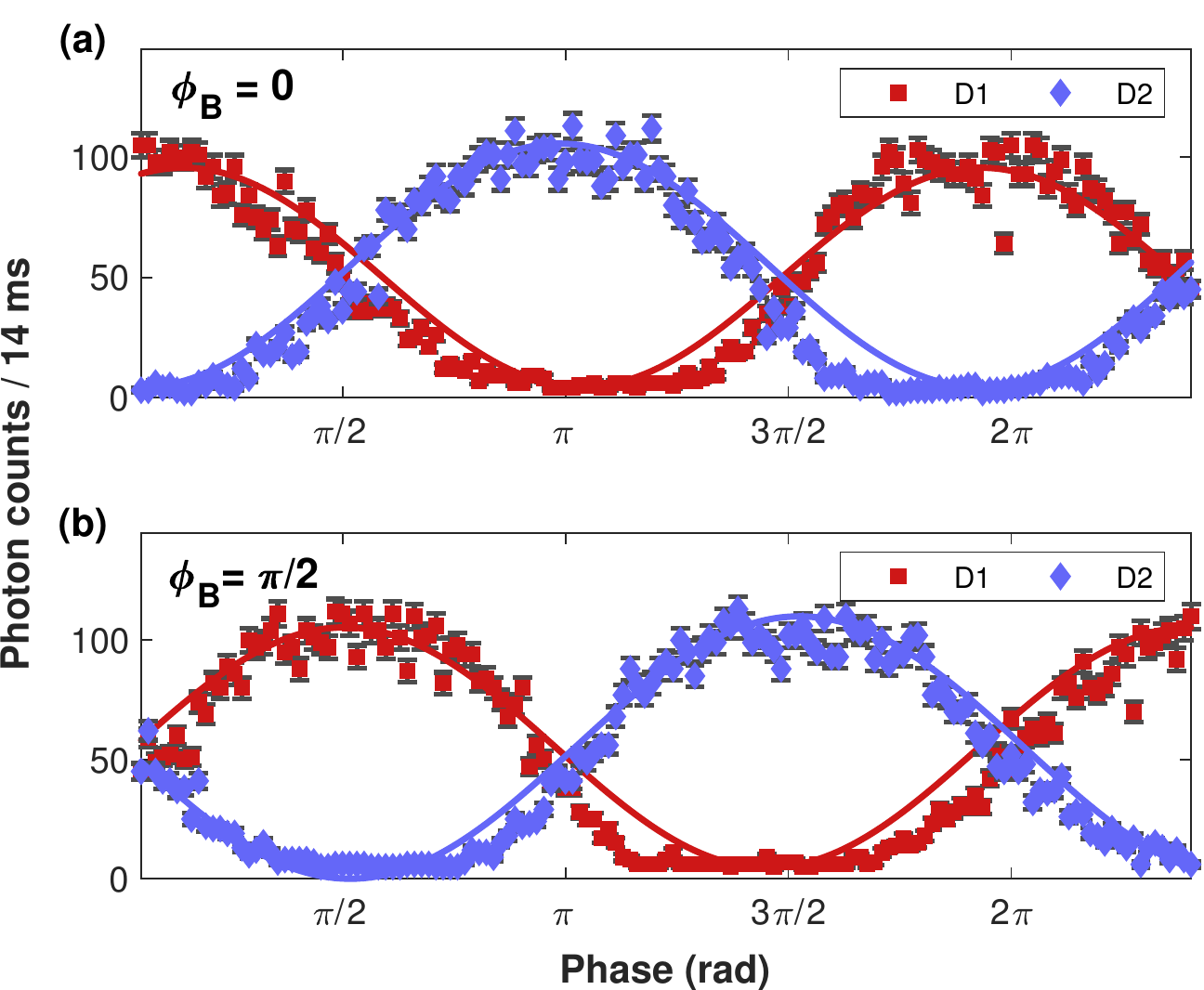}
\caption{Single-photon interference curves after 500 m propagation over the few-mode fiber. The relative phase $(\phi_A)$ is modulated with a slow-driving triangular signal. a) Interference curves recorded while the detection stage is set for measuring at the first mutually unbiased base defined by $\phi_B = 0$. b) Interference curves recorded while the detection stage is set for measuring at the second mutually unbiased base defined by $\phi_B = \pi/2$. The standard deviation is calculated assuming Poissonian statistics. The deviation that can be seen from the experimental data to the theoretical fit is due to the residual phase drift present acting on Alice and Bob's local interferometers that are not actively phase-stabilized. \label{Fig3} }
\end{figure}

In order to demonstrate the feasibility of quantum communication protocols \cite{Xu_2020, Pirandola_2020} in our setup, we removed the CCD camera, reconnected the variable optical attenuator, and connected the output of the FMF to Bob's station as shown in Fig. \ref{Fig1}c. The required states $|\mathrm{LP}_{\pm}\rangle$ and $|\mathrm{OAM}_{\pm}\rangle$ are again prepared by driving $\phi_A$ sequentially. The variable attenuator is set to create a weak coherent state with an average mean photon number of $\mu=0.4$ per gate width at Alice's output. Therefore, contribution of multi-photon events is negligible in our experiment. We acquire the detection counts with a short integration time (14 ms) to visualise the interference pattern without active phase stabilisation of the first local interferometer. The single-photon interference patterns obtained after 500 m of propagation over the FMF are shown in Fig. \ref{Fig3}. Figure \ref{Fig3}a is the case corresponding to the detection at the first mutually unbiased base defined by $\phi_B = 0$. In Fig. \ref{Fig3}b, we have the single-photon interference curves obtained with the detection apparatus operating at the second MUB defined by $\phi_B = \pi/2$.

\begin{figure}[ht!]
\centering\includegraphics[width=\columnwidth]{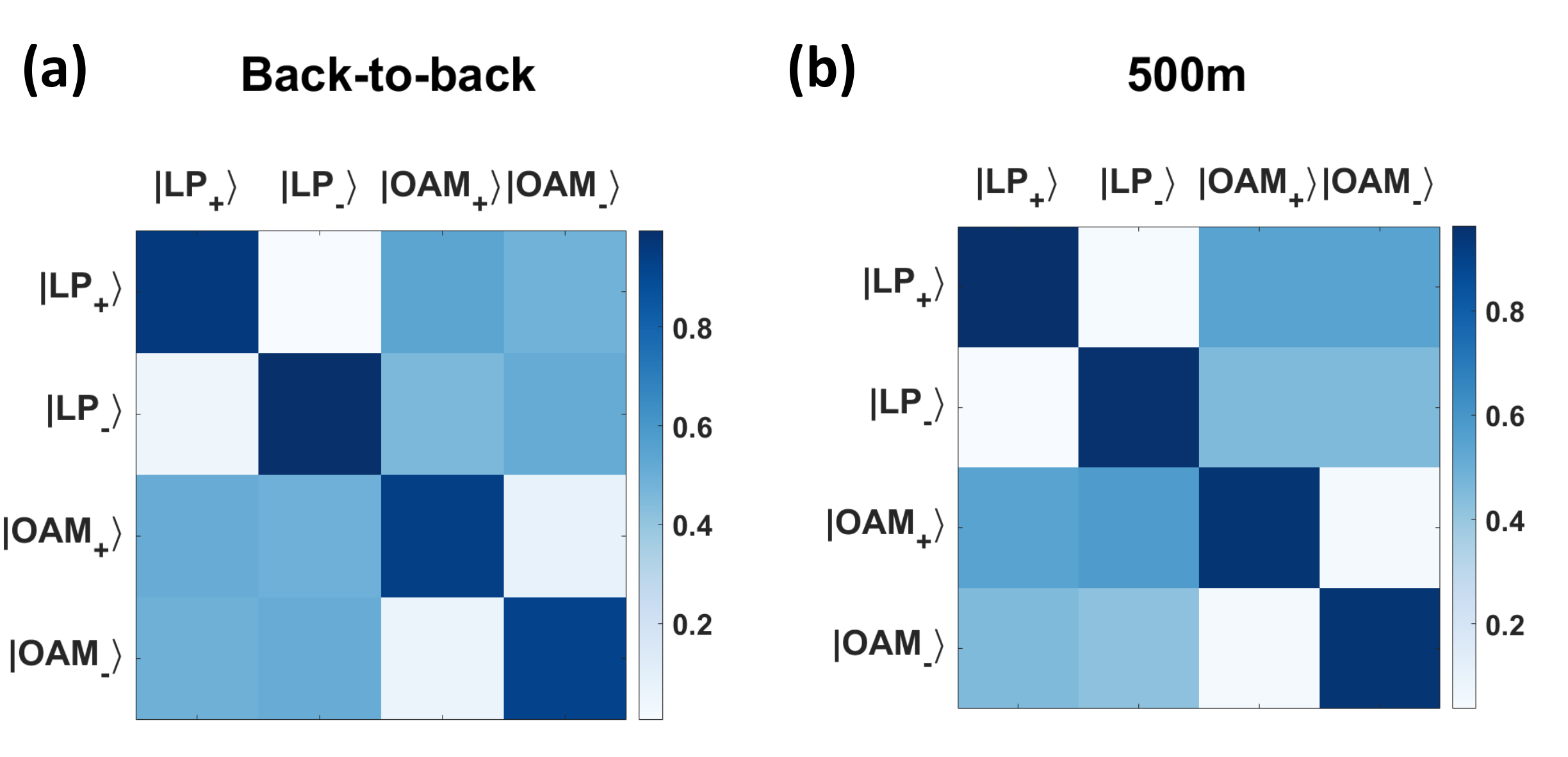}
\caption{Probabilities associated to the states of the BB84 quantum cryptography protocol. The vertical axis show the states being prepared, while the horizontal one shows to which state the projection is made onto. a) The probabilities in the back-to-back case, while b) shows the same probabilities following 500 m of propagation over the FMF. Please see the text for details. \label{Fig4} }
\end{figure}

From the data points displayed in Fig. \ref{Fig3} we calculate the probabilities of the projection onto state $i$ given that $j$ was sent, where $i, j = \{\ket{\mathrm{LP}_+}, \ket{\mathrm{LP}_-},\ket{\mathrm{OAM}_+}, \ket{\mathrm{OAM}_-}\}$.  The results for the transmitted vs. the projected states are plotted in Fig. \ref{Fig4} for both the back-to-back and 500 m cases. The average probability for the main diagonal (corresponding to the cases where the same state is being transmitted and projected onto) is $0.955 \pm 0.022$ and $0.951 \pm 0.024$ for 0 and 500 m respectively. This points to a lower bound in the error rate for quantum key distribution using this setup and employing the BB84 protocol of less than 5\% at 500 m, showing the feasibility of this scheme. Out of that number, 0.016\% is due to dark counts of the detector. Also, as expected in BB84 protocol, the probabilities when performing measurements in the non-matching bases is around 50\%. Furthermore, by considering a simple attenuation model in the single-photon interference curves, we calculate that an extra 3.85 dB loss (corresponding to 17.5 kms of FMF)  is needed to bring the QKD error rate to 11\%, a higher bound where usually no secret key can be generated.

\section{Conclusion}
We have proposed and experimentally demonstrated a phase-encoded quantum communication system based on few-mode fibers, which removes the irreversible detection loss that is present on all time-bin quantum communication systems. While it would be possible to employ the polarization degree-of-freedom to implement this idea, i.e. sending orthogonal polarization states in the same fiber, it would not allow an expansion to higher dimensions. Our scheme on the other hand can be further upgraded to higher dimensions by using few-mode fibers \cite{Sillard_2014}, lanterns \cite{Velazquez_2018} and beamsplitters \cite{Carine_2020} supporting more modes, while still not presenting extra intrinsic detection losses, thus solving this fundamental issue regarding high-dimensional time-bin encoding (more details in the Appendix). Furthermore, our proposal becomes even more effective in a $d$-dimensional space, due to the growth of the loss with $(d-1)/d$. Recently multi-core fibers (MCFs) have been successfully used to transport spatially encoded quantum states \cite{Canas_2017}, but they still suffer from a slow phase drift due to the fact each mode takes a separate core in the fiber \cite{DaLio_2019}. FMFs on the other hand show no such drift, since the modes propagate in the same core (See the Appendix), showing they may be ideal for this application. Other recently demonstrated lanterns have had losses as low as 0.7 dB for a 6-mode lantern \cite{Velazquez_2015}, with recent simulations pointing out that lanterns of much lower losses (0.1 dB) could be reached \cite{Li_2019}, further demonstrating the attractiveness of our scheme as a path in the future use of time-bin encoding for quantum communication. For the specific case of qubit-based time-bin quantum communication systems, if one employs improved lanterns with 0.7 dB losses, then there is an increase in approximately 70\% in the overall detection efficiency compared to a standard time-bin system, with the gain increasing if higher-dimensional systems are employed. Finally, although we used BB84 as an illustrative example of the possibilities of our setup for quantum information, other applications can benefit directly from the removed detection losses such as high-dimensional quantum cryptography \cite{Islam_2017b} and quantum random access codes \cite{Tavakoli_2015}.

Another major achievement, that further shows that SDM technology can provide more significant benefits to quantum information \cite{Xavier_2020}, is the use of photonic lanterns to generate and decode OAM spatial states completely in-fiber, which also has important applications in optical communications \cite{Bozinovic_2013}. Finally, recent developments in integrated photonic circuits \cite{Wang_2020}, could completely replace Alice and Bob's optical setups with integrated chips, greatly increasing compactness and robustness, and thus dismissing the need for active phase stabilisation. We therefore envisage these results will have significant and an imminent impact in areas such as long-distance quantum communication, high-dimensional quantum information and as a tool to further increase the capacity of classical communication networks.

\section{Appendix}
\subsection{Expansion to higher dimensions}

\begin{figure}[ht!]
\centering\includegraphics[width=\columnwidth]{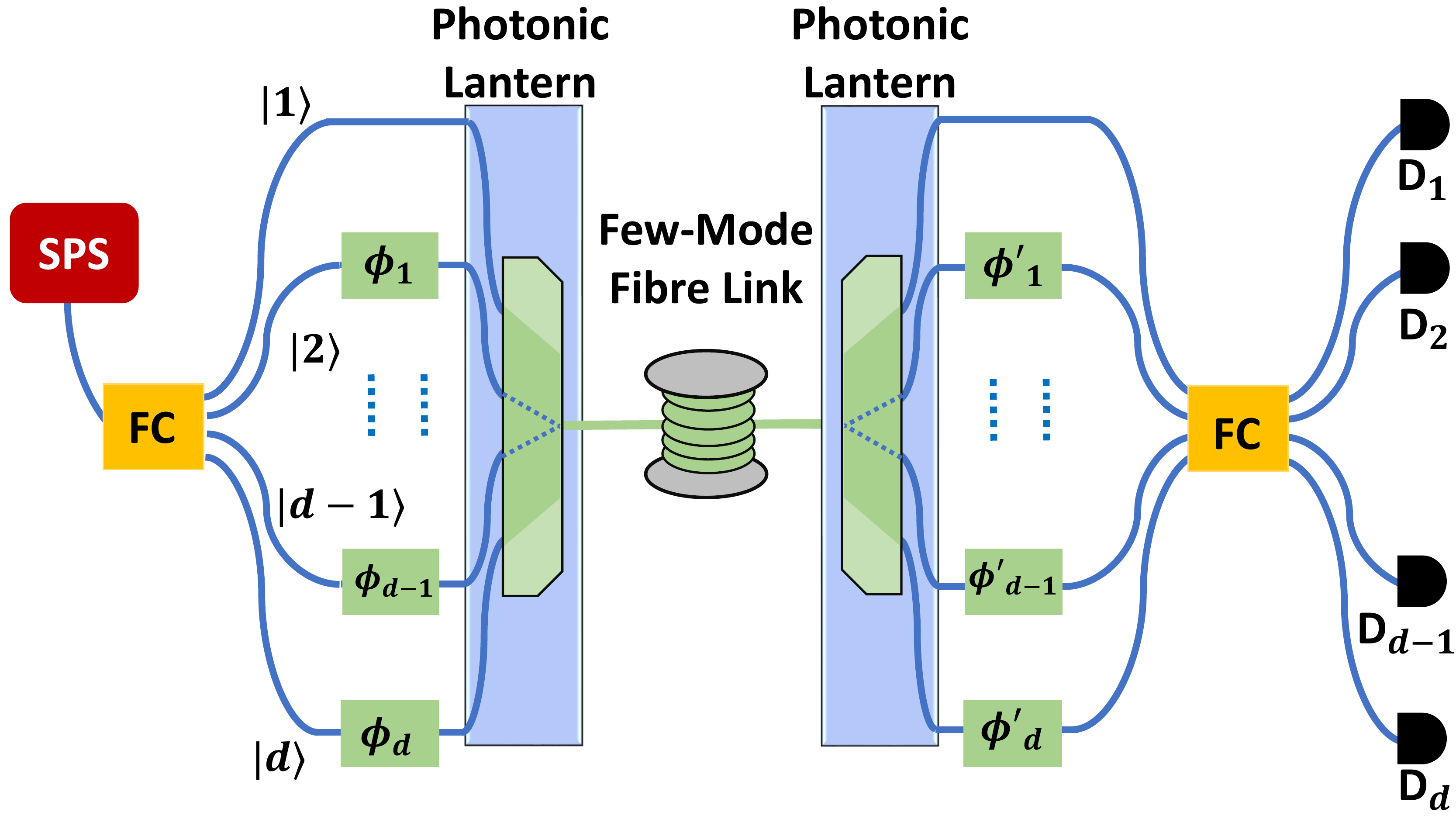}
\caption {Qudit-based quantum communication scheme using phase-encoding with few-mode fibers.}   \label{Fig5}
\end{figure}

Our scheme is directly scalable by employing SDM components that support a higher number of modes: lanterns \cite{Velazquez_2018}, few-mode fibers \cite{Sillard_2014} and multi-port beamsplitters \cite{Carine_2020}. Furthermore, our scheme becomes even more beneficial at higher dimensions, due to the greatly increasing loss with $d$ for the time-bin configuration. The proposal is shown in Fig. \ref{Fig5}, where we split the output of the single-photon source into a total of $d$ paths, using a multi-port beamsplitter (MBS).  Each path has a different phase modulator $\phi_n$, with $n$ ranging from 0 to $d-1$, allowing the preparation of the high-dimensional path state $|\Psi\rangle = 1/\sqrt{d}\sum_{n=0}^{d-1}e^{i\phi_n}|n\rangle$, where $|n\rangle$ represents the $n$th path and $1/\sqrt{d}$ is a normalization factor. Then each path is connected to a $d$-mode photonic lantern in order to excite the corresponding LP mode in a $d$-mode FMF. Finally, another $d$-mode lantern is used to demultiplex the LP modes into the corresponding $d$ paths, which contain the phase modulators $\phi_n'$ to choose the state projection. Finally, an $d \times d$ MBS is employed to superpose the different paths, and the $d$ outputs are connected to single-photon detectors. Please note that no temporal post-selection is needed, as in the bi-dimensional case.

\subsection{Phase stability of phase-encoded states over a FMF}

As discussed in the main text, phase encoding is not usually employed in the original Mach-Zehnder configuration since the channel is not held stable long enough for a key exchange to take place, due to fast phase instabilities. Although MCFs show a considerable improvement in this aspect when compared to independent single-mode fibers, they still have a residual drift which requires active phase drift compensation \cite{Canas_2017, DaLio_2019}. FMFs on the other hand, have the advantage that the individual spatial modes travel through the same core, which is more intrinsically stable than independent cores in the same cladding.

\begin{figure*}[ht!]
\centering\includegraphics[width=16cm]{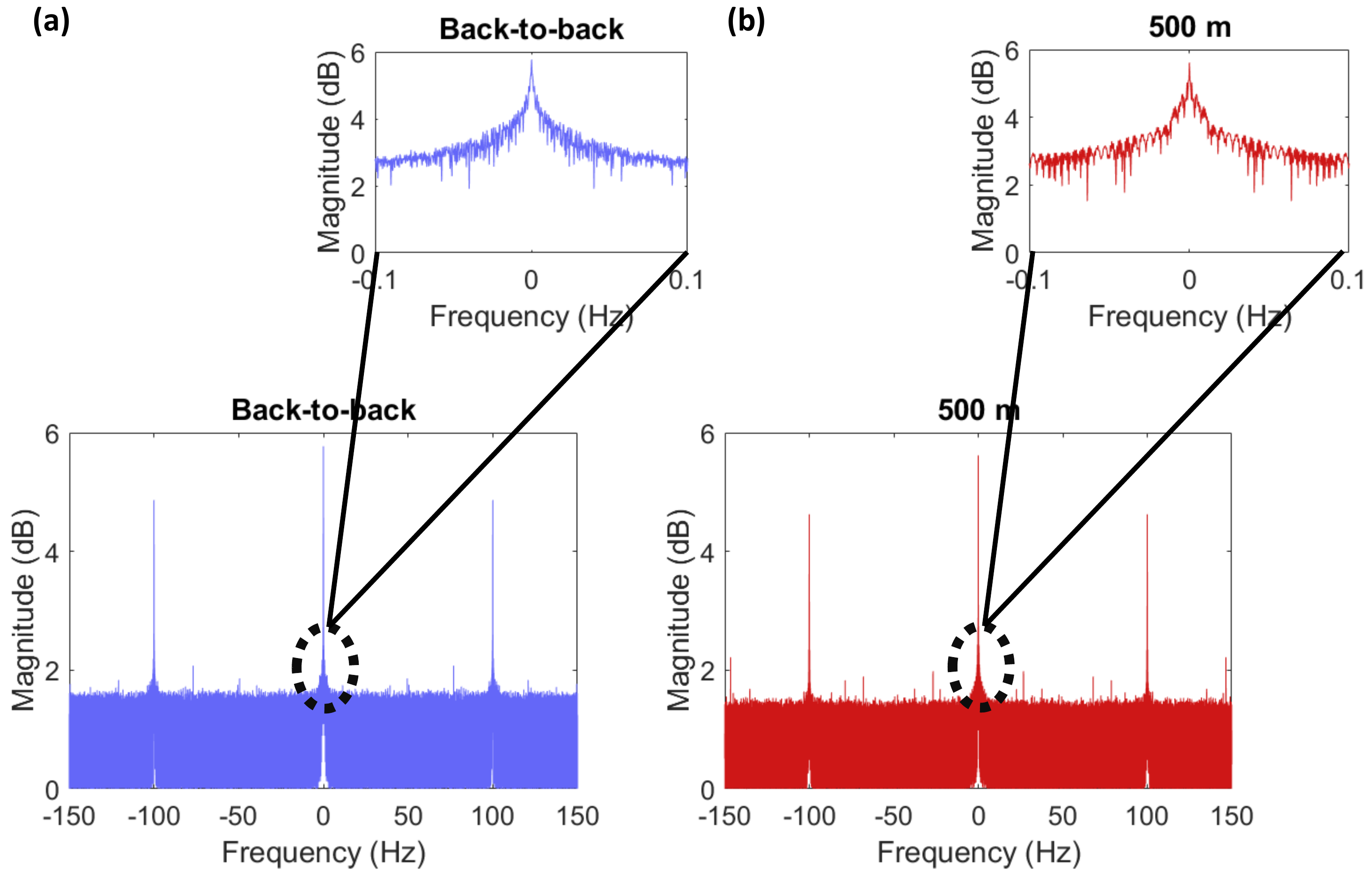}
\caption {Fourier spectra of 50 min-long measurements of the interferometer output, when the input state is continuously modulated with a 100 Hz sinusoidal wave, for both the back-to-back (a) and after an extra 500 m of FMF (b).}   \label{Fig6}
\end{figure*}

In order to demonstrate the stability, we  adjusted the attenuator at the output of the laser source to work with classical optical power levels. We then continuously modulated the input state through the phase modulator $\phi_A$ with a 100 Hz sinusoidal wave during 50 mins, both in the back-to-back case (10 m), and with the 500 m long spool connected. The output from the final beamsplitter in Fig. \ref{Fig1}c was measured with a p-i-n photodiode. The driving voltage is meant to impose a clear modulation signal creating well-defined interference fringes. The output from the photodiode is recorded with an oscilloscope. Figures \ref{Fig6}a and \ref{Fig6}b shows the Fourier transforms of the recorded 50 minutes time signal for the back-to-back and 500 m cases respectively. The inset shows zooms in around the centre peak at 0 Hz. The environmental phase drift acting on an interferometric setup in the lab is typically characterised by low frequency components. In our case, there is substantial phase drift acting over Alice and Bob's components, which consist of parallel paths (single-mode fibers). Comparing both spectra (back-to-back and 500 m) it is clear there is no significant difference in the low frequency region, or even across a broader frequency range. This result points out to the fact that adding an extra 500 m of FMF does not increase the environmental phase drift, which leads to the possibilities that FMFs can be used as a platform for phase-encoded states over longer distances.

To further corroborate the benefits of using a FMF in this regard, we perform a brief comparison based on two multi-core fiber interferometers previously used in other experiments. In \cite{Canas_2017}, it was demonstrated that tens of minutes were needed for the environmental phase drift to cause a complete inversion in the detected state in a 300 m long multi-core fiber interferometer. In \cite{DaLio_2019} it can be seen that an equivalent measurement on a 2-km long multi-core fiber interferometer is on the order of tens of seconds. While it is difficult to make a precise measurement on the difference, due to the fact the two experiments were done in different labs, it still shows a clear difference when increasing the distance from 300 m to 2 kms in multi-core fiber interferometers. In our case, as we can observe no appreciable difference when moving from back-to-back to 500 m, this further shows the benefits of using FMFs in this regard.

\section*{Acknowledgments}
We acknowledge Ceniit Link\"{o}ping University, the Swedish Research Council (VR 2017-04470), QuantERA grant SECRET (VR grant no. 2019-00392) and the Knut and Alice Wallenberg Foundation through the Wallenberg Center for Quantum Technology (WACQT) for financial support. G.L. was supported by Fondo Nacional de Desarrollo Cient\'{i}fico y Tecnol\'{o}gico (FONDECYT) (1200859) and ANID - Millennium Science Initiative Program - ICN17\_012).


\end{document}